\documentclass[conference]{IEEEtran}
\IEEEoverridecommandlockouts

\usepackage{cite}
\usepackage{amsmath,amssymb,amsfonts}
\usepackage{algorithmicx}
\usepackage{graphicx}
\usepackage{textcomp}
\usepackage{xcolor}
\usepackage{comment}
\usepackage[acronym]{glossaries}
\usepackage{changepage}
\usepackage{threeparttable}
\usepackage{subcaption}
\usepackage{pgfplotstable}
\usepackage{pgfplots}
\usepackage{tikz}
\usepackage{setspace} 
\usepackage{siunitx}
\usepackage{algorithm}
\usepackage[noend]{algpseudocode}
\usepackage{MnSymbol,wasysym}
\usepackage{cleveref}
\usepackage{booktabs}
\usepackage{multirow}
\usepackage{url}
\usepackage[skip=2pt, font=small]{caption}

\Crefname{equation}{Eq.}{Eqs.}
\Crefname{figure}{Fig.}{Figs.}
\Crefname{tabular}{Tab.}{Tabs.}
\Crefname{algorithm}{Alg.}{Algs.}

\newcommand\modified[1]{\textcolor{black}{#1}}
\newcommand\modifiednew[1]{\textcolor{black}{#1}}

\def\BibTeX{{\rm B\kern-.05em{\sc i\kern-.025em b}\kern-.08em
    T\kern-.1667em\lower.7ex\hbox{E}\kern-.125emX}}

\usetikzlibrary{shapes, arrows, patterns, calc}
\tikzset{>=latex}

\tikzstyle{block} = [draw, thick, rectangle, minimum height=0.75cm, minimum width=0.75cm]
\tikzstyle{sum} = [draw, fill=white, circle, node distance=1cm, thick]
\tikzstyle{gain} = [
  regular polygon,
  regular polygon sides=3,
  draw,
  thick,
  fill=white,
  text width=0.3em,
  inner sep=0.3mm,
  outer sep=0mm,
  shape border rotate=-90
]
\tikzstyle{revgain} = [
  regular polygon,
  regular polygon sides=3,
  regular polygon rotate=90,
  draw,
  thick,
  fill=white,
  text width=0.3em,
  inner sep=0.3mm,
  outer sep=0mm,
]
\usetikzlibrary{arrows.meta}
\tikzset{myarrow/.style={>={Computer Modern Rightarrow[scale length=1.5]}}}

\pgfplotsset{compat=1.8}
\pgfplotsset{compat=1.16}
\definecolor{colorblue}{rgb}{0.12156862745098,0.466666666666667,0.705882352941177} 
\definecolor{colorgreen}{rgb}{0.172549019607843,0.627450980392157,0.172549019607843} 
\definecolor{colorred}{rgb}{0.83921568627451,0.152941176470588,0.156862745098039} 
\definecolor{newblue}{RGB}{0,113,188}
\definecolor{newred}{RGB}{217,82,24}
\definecolor{darkred}{RGB}{140,0,0}
\definecolor{darkerred}{RGB}{106,46,36}
\definecolor{bargrey}{RGB}{165,165,165}
\definecolor{bardarkblue}{RGB}{68,114,196}
\definecolor{barblue}{RGB}{91,155,213}
\definecolor{bargreen}{RGB}{112,173,71}
\definecolor{baryellow}{RGB}{255,192,0}
\definecolor{barred}{RGB}{237,125,49}
\pgfplotsset{every axis/.append style={
font=\footnotesize,
label style={font=\footnotesize},
tick label style={font=\scriptsize}  
}
}

\makeatletter
\def\ps@IEEEtitlepagestyle{%
  \def\@oddfoot{\mycopyrightnotice}%
  \def\@oddhead{\hbox{}\@IEEEheaderstyle\leftmark\hfil\thepage}\relax
  \def\@evenhead{\@IEEEheaderstyle\thepage\hfil\leftmark\hbox{}}\relax
  \def\@evenfoot{}%
}
\def\mycopyrightnotice{%
  \begin{minipage}{\textwidth}
  \centering \scriptsize
  \copyright 2024 IEEE.  Personal use of this material is permitted.  Permission from IEEE must be obtained for all other uses, in any current or future media, including reprinting/republishing this material for advertising or promotional purposes, creating new collective works, for resale or redistribution to servers or lists, or reuse of any copyrighted component of this work in other works.
  \end{minipage}
}
\makeatother
    
\begin{document}

\title{Train-On-Request: An On-Device Continual Learning Workflow for Adaptive Real-World Brain Machine Interfaces
}

 \author{\IEEEauthorblockN{
    Lan Mei\IEEEauthorrefmark{2},
    Cristian Cioflan\IEEEauthorrefmark{2},
    Thorir Mar Ingolfsson\IEEEauthorrefmark{2},
    Victor Kartsch\IEEEauthorrefmark{2},\\
    Andrea Cossettini\IEEEauthorrefmark{2},
    Xiaying Wang\IEEEauthorrefmark{1}\IEEEauthorrefmark{2},
    Luca Benini\IEEEauthorrefmark{2}\IEEEauthorrefmark{3}
    }
    \IEEEauthorblockA{\\\IEEEauthorrefmark{2}Dept. ITET, ETH Zürich, Zürich, Switzerland \hspace{3.2mm}\IEEEauthorrefmark{3}DEI, University of Bologna, Bologna, Italy\vspace{-0.2cm}}
    
    \thanks{* Corresponding author: xiaywang@iis.ee.ethz.ch}
    \thanks{We open-source release the codes: https://github.com/pulp-bio/bmi-odcl.git}
    
    \vspace{-0.5cm}
    }

\maketitle

\begin{abstract}

Brain-machine interfaces (BMIs) are expanding beyond clinical settings thanks to advances in hardware and algorithms. However, they still face challenges in user-friendliness and signal variability. Classification models need periodic adaptation for real-life use, making an optimal re-training strategy essential to maximize user acceptance and maintain high performance. We propose TOR, a train-on-request workflow that enables user-specific model adaptation to novel conditions, addressing signal variability over time. Using continual learning, TOR preserves knowledge across sessions and mitigates inter-session variability. With TOR, users can refine, on demand, the model through on-device learning (ODL) to enhance accuracy adapting to changing conditions. We evaluate the proposed methodology on a motor-movement dataset recorded with a non-stigmatizing wearable BMI headband, achieving up to 92\% accuracy and a re-calibration time as low as 1.6 minutes, a 46\% reduction compared to a naive transfer learning workflow. We additionally demonstrate that TOR is suitable for ODL in extreme edge settings by deploying the training procedure on a RISC-V ultra-low-power SoC (GAP9), resulting in 21.6 ms of latency and 1 mJ of energy consumption per training step. To the best of our knowledge, this work is the first demonstration of an online, energy-efficient, dynamic adaptation of a BMI model to the intrinsic variability of EEG signals in real-time settings.

\end{abstract}

\begin{IEEEkeywords}
 brain-machine interface, EEG, wearable EEG, wearable healthcare, transfer learning, continual learning, on-device learning
\end{IEEEkeywords}

\newacronym{eeg}{EEG}{electroencephalography}
\newacronym{bmi}{BMI}{Brain--Machine Interface}
\newacronym{soa}{SoA}{State of the Art}
\newacronym{dl}{DL}{Deep Learning}
\newacronym{cl}{CL}{Continual Learning}
\newacronym{tl}{TL}{Transfer Learning}
\newacronym{odl}{ODL}{On-Device Learning}
\newacronym{lwf}{LwF}{Learning without Forgetting}
\newacronym{er}{ER}{Experience Replay}
\newacronym{cnn}{CNN}{Convolutional Neural Network}
\newacronym{snr}{SNR}{signal-to-noise ratio}
\newacronym{tor}{TOR}{Train-on-Request}
\newacronym{mm}{MM}{Motor Movement}
\newacronym{mi}{MI}{Motor Imagery}
\newacronym{tcdm}{TCDM}{Tightly-Coupled Data Memory}
\newacronym{fc}{FC}{Fabric Controller}
\newacronym{dnn}{DNN}{Deep Neural Network}
\newacronym{mcu}{MCU}{Microcontroller Unit}
\newacronym{fpu}{FPU}{Floating Point Unit}
\newacronym{simd}{SIMD}{Single Instruction Multiple Data}
\newacronym{itr}{ITR}{Information Transfer Rate}

\section{Introduction}
\label{sec:introduction}

\Glspl{bmi} bridge the communication gap between users and devices by utilizing brain signals such as \gls{eeg}. 
Beyond the usage in laboratories and hospitals, the recent development of consumer-grade \gls{bmi} \modified{is targeting} a wider range of markets~\cite{emotiv,muse}.
A common category of \gls{bmi} applications employs \gls{eeg} signals recorded while subjects use their motor functions, i.e., \gls{mm}, or imagine the movements, i.e., \gls{mi}, of body parts such as hands, feet, or tongue~\cite{pfurtscheller2001functional, morash2008classifying}.
These approaches \modified{could} enable a wide range of \gls{bmi} applications, such as automatic gaming~\cite{malete2019eeg} and rehabilitative robotics~\cite{lopez2018brain}.

Such systems commonly rely on pre-trained models~\cite{malete2019eeg}, yet exposure to novel usage conditions (i.e., sessions) or novel subjects leads to performance degradation over time~\cite{herff2020potential}.
Current \gls{bmi} systems lack hardware and learning mechanisms to learn from the dynamic nature of~\gls{eeg} signals and train the models at the edge.
Nevertheless, adapting to the user-specific, on-site conditions is crucial for intelligent systems in order to preserve their accuracy~\cite{di2023personalized, mukherjee2024personalization}. 
Specifically, real-time, on-device adaptation is necessary to mitigate inter-session variability for \gls{eeg}-based \gls{bmi} systems~\cite{wang2023enhancing, pyun2024machine}.

Recently, Wang et al.~\cite{wang2023enhancing} proposed a chain-\gls{tl} framework that enables online finetuning with an initial calibration phase. 
A model is initially pretrained on the first session acquired for the target subject.
For each subsequent session, the user is requested to record a set of training data at the beginning of the session on which the model is finetuned. Afterwards, the accuracy is evaluated on the following, remaining data of the session.
While such \gls{tl} finetuning strategy has been shown to increase the accuracy in novel, unseen conditions~\cite{cioflan2024device}, the model forgets the variability of the data seen previously, i.e., the accuracy on previously learned conditions decreases over time, a phenomenon known as catastrophic forgetting~\cite{nguyen2019toward}.
\gls{cl} techniques further improve long-term classification performance by mitigating catastrophic forgetting issues, yet they are still to be fully explored for \gls{eeg}-based \gls{bmi} applications. 

An additional challenge in this context is the lengthy data acquisition process, which can cause boredom and drowsiness~\cite{penaloza2014brain} in the subjects, as well as the tedious labeling procedure that affects user experience. 
Nonetheless, it is crucial to have sufficient \gls{eeg} recordings, as the performance of \gls{dl}-based \gls{bmi} depends significantly on the amount of available data~\cite{al2021deep, hossain2023status}. 
Moreover, the complexity and noise of in-the-wild \gls{eeg} signals cause low \gls{snr} levels~\cite{pijn1991chaos} and further prolong the setup and data acquisition process.
It is thus important to balance reasonable data acquisition time and satisfactory performance when designing \gls{bmi} systems, namely, new training acquisitions should be performed only when really needed.

To address these challenges, we present a fine-grained adaptation methodology suitable for extreme-edge \gls{bmi} systems.
We denote our workflow as \gls{tor}, as it offers the user complete control in deciding when to initiate a learning process.
The main contributions of this work are as follows:

\begin{itemize}
\item We propose a novel, flexible \gls{tor} workflow capable of online adaptations based on classification performance, \modified{improving} the user experience through near real-time data acquisition;
\item 
We introduce \gls{cl}-based \gls{tor} to further mitigate inter-session variability. 
In particular, \gls{er}-based \gls{tor} achieves up to 92.33\% accuracy while requiring 46.67\% less training data than chain-\gls{tl} workflows~\cite{wang2023enhancing}, reducing the acquisition time by 28 minutes;
\item We demonstrate the suitability of \gls{tor} for edge processors, by showing improved classification accuracy in only \qty{21.6}{\milli \second} per training step and an energy consumption as low as \qty{1}{\milli \joule} on the GAP9 \gls{mcu}. 
\end{itemize}

\section{Materials and Methods}
\label{sec:materialsandmethods}

\subsection{Dataset, Preprocessing, and Classification Model}
\label{ssc:datamanagement}

In this work, we use the dataset introduced in~\cite{wang2023enhancing} consisting of EEG data collected from a healthy subject over seven sessions using a wearable \gls{bmi} headband based on BioWolf~\cite{kartsch2019biowolf} at a sampling rate of~\qty{500}{\hertz}.
Following instructions shown on a screen, the subject performs \gls{mm} tasks of left- and right-hand finger-tapping~\cite{chang2018_bcitutorial}. 
The instruction is shown for~\qty{4}{\second} on the screen, followed by a resting period of five to six seconds.
We use all seven sessions of \gls{eeg} data and take from each the first ten runs, each with five trials of left-hand movement and five of right-hand movement, to ensure experimental consistency, resulting in 100 trials per session for two-class \modified{ (i.e., left- and right-hand \gls{mm})} classification.

We preprocess the raw \gls{eeg} data with a 4$^{th}$ order band-pass filter of~\qty{0.5}-\qty{100}{\hertz}, a notch filter to suppress~\qty{50}{\hertz} power line interference, and a moving average filter with a sliding window of \qty{0.25}{\second}.
We use MI-BMInet~\cite{wang2024mi}, the \gls{soa} lightweight \gls{cnn} for \glspl{bmi} on ultra-low-power edge devices, shown in Table~\ref{tab:MIBMINetStructureTable}. 

\begin{table}[!b]
 \caption{MI-BMINet~\cite{wang2024mi} architecture with depthwise (DW), pointwise (PW), and depthwise separable (DS) convolutions.}
 \label{tab:MIBMINetStructureTable}
  \centering
 \resizebox{\linewidth}{!}{
 \setlength{\tabcolsep}{4pt}
\begin{tabular}{@{}llll@{}} \toprule
  \textbf{Blocks} & \textbf{Layers} & \textbf{\# Filters} & \textbf{Size} \\ 
  \midrule
  Spatial Conv. & DW Conv. + BN & $N_k=32$ & $8 \times 1$ \\
  Temporal Conv. & DW Conv. + BN + ReLU + Pool. & $N_k=32$ & $1 \times 128$ \\
  \multirow{2}{*}{DS Conv.} & DW Conv. & $N_k=32$ & $1 \times 16$ \\
   & PW Conv. + BN + ReLU + Pool. & $N_k=32$ & $1 \times 1$ \\ 
   Dense & Dense & $2 \times N_k \times (1900 // 64 )$  & - \\
   \bottomrule  
 \end{tabular} }
\end{table}

\subsection{``Train-On-Request" Workflow}
\label{ssc:tor}
\begin{figure}[!t]
  \centering
  \input ./figures/TOR_workflow.tex
\end{figure} 

\begin{algorithm}[b]
\caption{\gls{tor} workflow in novel sessions.}
\label{alg:tor}
\begin{algorithmic}
\For{$subs=1$ to $subss$}
    \State Test on subsession $subs$ 
    \If{$Acc \geq T_{Acc}$}
        \State \textbf{continue}
    \Else
    \State $subs \gets subs+1$
    \For{$ep=1$ to $eps$}
        \For {$trl$ in $trls$} 
            \State Finetune on trial $trl$ of subsession $subs$
        \EndFor
    \EndFor
    \EndIf
\EndFor

\end{algorithmic}
\end{algorithm}

We propose the \gls{tor} workflow where the model is adaptively updated based on user experience. 
An example of the workflow is illustrated in~\Cref{fig:TOR_workflow}. 
Considering different experimental sessions with varying conditions, \gls{tor} enables fine-grained evaluation within each session.
When the accuracy is deemed unsatisfactory (e.g., more than one failed command out of ten attempts in controlling a \gls{bmi}), the user activates \gls{tor}, which triggers a retraining phase where the model adapts to the novel condition.

Detailed implementation of \gls{tor} within a new session is described in~\Cref{alg:tor}. 
Similar to chain-\gls{tl}~\cite{wang2023enhancing}, we pretrain our model on the first data session as a starting point of the workflow. 
The streaming data of each new session, which contains $100$ trials in our dataset, is divided into $subss=10$ subsessions (SS) of $trls=10$ trials each. 
An accuracy threshold $T_{Acc}=90\%$ models the user satisfaction with the network performance and dictates when additional learning is needed.
Starting from SS$_1$ of a novel session, the existing model is evaluated for every incoming subsession.
Once an unsatisfactory accuracy level is measured on SS$_{i}$, all trials of the SS$_{i+1}$ are used to finetune the model for $eps$ epochs. 
The evaluation process is then performed on SS$_{i+2}$, and the user can exploit the updated model until a new finetuning is requested (or the session is complete).

Similarly to the chain-\gls{tl} methodology~\cite{wang2023enhancing}, the pre-trained model is obtained by training on first session data for 40 epochs with a learning rate of $10^{-3}$. 
Then, we sequentially adapt the model for each of the next new sessions (i.e., six in this work), training our model for 15 epochs per subsession with a learning rate of $2\times10^{-3}$. 
\modifiednew{We implement chain-\gls{tl} as baseline following~\cite{wang2023enhancing} and the \gls{tor}-\gls{tl} based on \gls{tor} workflow.}
 
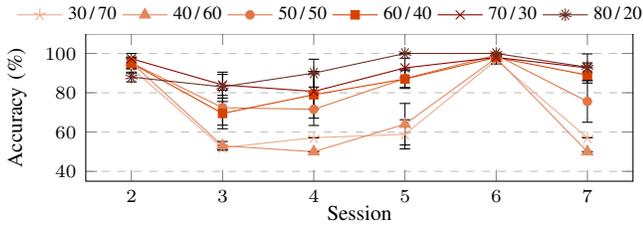
\begin{figure}[t]
\centering
    \begin{tikzpicture} 
    \begin{axis}[
        width=\linewidth,
        height=0.4\linewidth,
        xlabel={Session},
        ylabel={Accuracy (\%)},
        xmin=1.5, xmax=7.5,
        ymin=35, ymax=110,
        xtick={2,3,4,5,6,7},
        ytick={20,40,60,80,100},
        x label style={at={(axis description cs:0.5,-0.1)},anchor=north},
        legend cell align={left},
        legend columns = 6,
        legend style={at={(0.45,1.26)},anchor=north,draw=none, font=\scriptsize},
        ymajorgrids=true,
        grid style=dashed,
    ]
    
    \addplot[
        color=newred!50,
        mark=star, 
        mark options={scale=1.2},
        error bars/.cd, y dir=both, y explicit, error bar style={line width=0.25pt, color=black, solid}
        ]
        coordinates {
        (2,92.57) +- (0,2.77)
        (3,52.00) +- (0,1.14)
        (4,57.14) +- (0,0)
        (5,58.86) +- (0,7.42)
        (6,96.57) +- (0,1.94)
        (7,57.14) +- (0,0)  
        };
        \addlegendentry{\scriptsize 30\,/\,70}
    \addplot[
        color=newred!70,
        mark=triangle*,mark options={scale=1.2},
        error bars/.cd, y dir=both, y explicit, error bar style={line width=0.1pt, color=black, solid}
        ]
        coordinates {
        (2,95.33) +- (0,1.94)
        (3,53.00) +- (0,2.21)
        (4,50.00) +- (0,0.00)
        (5,64.00) +- (0,10.57)
        (6,97.67) +- (0,1.33)
        (7,50.00) +- (0,0.00) 
        };
        \addlegendentry{\scriptsize 40\,/\,60}
    \addplot[
        color=newred!80,
        mark=*,mark options={scale=0.8},
        error bars/.cd, y dir=both, y explicit, error bar style={line width=0.1pt, color=black, solid}
        ]
        coordinates {
        (2,95.20) +- (0,2.99)
        (3,72.40) +- (0,8.89)
        (4,71.60) +- (0,8.24)
        (5,87.20) +- (0,5.00)
        (6,99.20) +- (0,0.80)
        (7,75.60) +- (0,10.61)  
        };
        \addlegendentry{\scriptsize 50\,/\,50}
    \addplot[
        color=newred!110,
        mark=square*,mark options={scale=0.8},
        error bars/.cd, y dir=both, y explicit, error bar style={line width=0.1pt, color=black, solid}
        ]
        coordinates {
        (2,95.00) +- (0,2.74)
        (3,69.50) +- (0,7.81)
        (4,79.00) +- (0,11.90)
        (5,87.00) +- (0,4.30)
        (6,98.00) +- (0,1.00)
        (7,89.00) +- (0,4.06) 
        }; 
        \addlegendentry{\scriptsize 60\,/\,40} 
    \addplot[
        color=darkred,
        mark=x,mark options={scale=1.2},
        error bars/.cd, y dir=both, y explicit, error bar style={line width=0.1pt, color=black, solid}
        ]
        coordinates {
        (2,97.33) +- (0,2.67)
        (3,84.00) +- (0,5.33)
        (4,80.67) +- (0,9.04)
        (5,92.67) +- (0,4.90)
        (6,98.00) +- (0,1.63)
        (7,92.67) +- (0,2.49) 
        };
        \addlegendentry{\scriptsize 70\,/\,30}
    \addplot[
        color=darkerred,
        mark=10-pointed star,mark options={scale=1.2},
        error bars/.cd, y dir=both, y explicit, error bar style={line width=0.1pt, color=black, solid}
        ]
        coordinates {
        (2,88.00) +- (0,2.45)
        (3,83.00) +- (0,7.48)
        (4,90.00) +- (0,7.07)
        (5,100.00) +- (0,0.00)
        (6,100.00) +- (0,0.00)
        (7,93.00) +- (0,6.78) 
        };
        \addlegendentry{\scriptsize 80\,/\,20} 
    \end{axis}
    \end{tikzpicture}
\caption{Test accuracy over multiple sessions for chain-\gls{tl} baseline with different train/test data splits.} 
\label{fig:OptTrainTestSplit_Acc}
\vspace{-0.25cm}
\end{figure}

\subsection{Continual Learning Techniques} 
\label{ssc:continuallearning}

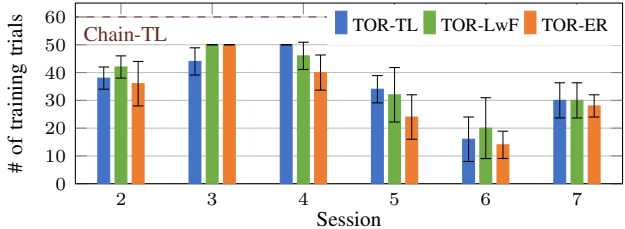
\begin{figure}[h!] 
\centering
\begin{tikzpicture} 
\begin{axis}[
	width=\linewidth,
    height=0.45\linewidth,
    xmin=1.5, xmax=7.5,
    ymin=0, ymax=65,
    xtick={2,3,4,5,6,7},
    ytick={0,10,20,30,40,50,60},
	xlabel=Session,
    ylabel=\# of training trials,
    x label style={at={(axis description cs:0.5,-0.1)},anchor=north},
	legend style={at={(0.999,0.99)},
	anchor=north east,legend columns=3, draw=none,
 font=\scriptsize},
	ybar,
    bar width=4.5pt,
    ymajorgrids=true,
    xtick align=inside,
    legend image code/.code={
        \draw [#1] (0cm,-0.1cm) rectangle (0.15cm,0.2cm); },
]
\addplot[
    color=bardarkblue, fill=bardarkblue, error bars/.cd, y dir=both, y explicit, error bar style={line width=0.25pt, color=black}
    ]
	coordinates {
    (2,38) +- (0,4)
    (3,44) +- (0,4.90)
	(4,50) +- (0,0)
    (5,34) +- (0,4.90)
    (6,16) +- (0,8)
	(7,30) +- (0,6.32)};
\addplot[
    color=bargreen, fill=bargreen, error bars/.cd, y dir=both, y explicit, error bar style={line width=0.25pt, color=black}
    ]
	coordinates {
    (2,42) +- (0,4)
    (3,50) +- (0,0)
	(4,46) +- (0,4.90)
    (5,32) +- (0,9.80)
    (6,20) +- (0,10.95)
	(7,30) +- (0,6.32)};
\addplot[
    color=barred, fill=barred, error bars/.cd, y dir=both, y explicit, error bar style={line width=0.25pt, color=black}
    ]
	coordinates {
    (2,36) +- (0,8)
    (3,50) +- (0,0)
	(4,40) +- (0,6.32)
    (5,24) +- (0,8)
    (6,14) +- (0,4.9)
	(7,28) +- (0,4)};
\addplot[darkerred, dashed, line legend,sharp plot,nodes near coords={},
update limits=false,shorten >=-3mm,shorten <=-3mm] 
    coordinates {(1,60) (8,60)}
    node[below,pos=0.15]{Chain-TL};
\legend{TOR-TL, TOR-LwF, TOR-ER}
\end{axis}
\end{tikzpicture}
\caption{Required training trials over multiple sessions for \gls{tl}- and \gls{cl}-based \gls{tor} workflows \modified{with $T_{Acc} = 90\%$, $trls=10$}. \modified{The horizontal dashed line represents the chain-\gls{tl} baseline.}} 
\vspace{-0.35cm}
\label{fig:NumOfTrainingTrials}
\end{figure}

To improve the accuracy in new sessions while preserving knowledge acquired from previous ones, 
\modifiednew{we explore \gls{cl}-based approaches in addition to \gls{tor}-\gls{tl}.} 
We compare the performance of two alternatives:
\begin{itemize}
\item \Gls{er}~\cite{rolnick2019experience}: uses a replay buffer to store previous samples, thus preserving the knowledge over past tasks and avoiding forgetting.
The replay buffer is updated in each session through \modified{reservoir sampling \cite{vitter1985random}}. 
This ensures that, in session $i$, the model is trained both with data from the current session, as well as with samples from all previous $i-1$ sessions.
In this work, we keep \gls{er} buffer size $s_{buf}=10$ in total throughout the workflow.

\item \Gls{lwf}~\cite{li2017learning}: uses the old model to compute the output given the new data.
The output is used as regularization term in the loss function, hereby preserving previous information during finetuning process. 
Compared to \cite{li2017learning}, we set the distillation hyperparameter $\lambda_o=1$ and temperature $T=2$. 
\end{itemize}
\modifiednew{We thus implement three adaptation strategies using \gls{tor} workflow: \gls{tor}-\gls{tl}, \gls{tor}-\gls{er}, and \gls{tor}-\gls{lwf}.
The chain-TL approach~\cite{wang2023enhancing} is also tested to provide a baseline.}

\section{Results}
\label{sec:results}

\subsection{Chain-\gls{tl} Baseline}

To establish a baseline, we first explore the optimal train-test split of the acquired dataset under the chain-\gls{tl} workflow~\cite{wang2023enhancing}.
\Cref{fig:OptTrainTestSplit_Acc} shows the chain-\gls{tl} test accuracy over six sessions, each averaged across five runs.
The measured accuracy generally increases with the number of training samples available, with the 80/20 split achieving the highest average accuracy of 92.33\% across six sessions. 
However, this split requires \qty{13.3}{\minute} of acquisition time for the training data \modified{for every session}.
To find an optimal trade-off between accuracy and training time, we note that a 60/40 split shows a satisfactory accuracy of 86.25\% while reducing the acquisition time to \qty{10}{\minute} per session.
These results allow us to identify a baseline for the next~\gls{tor} explorations, i.e., the chain-\gls{tl} with a 60/40 split of each 100 trials acquired during a session\modified{, treating the first 60 trials as training set and the last 40 trials as test set}.

\subsection{Transfer Learning and Continual Learning through \gls{tor}}

\pgfplotstableread{
  60   30
  20   20
  100  60
  100  20
  100  50
  90   20
  90   70
  100  20
  100  70
  100  20
}\datatable 

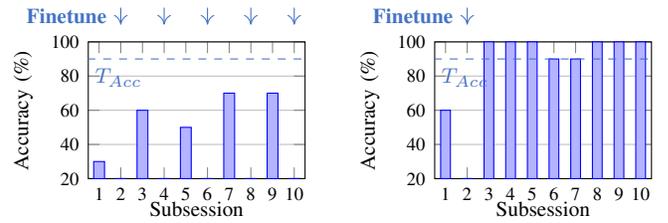
\begin{figure}[t]
 \centering
 \begin{subfigure}[t]{0.48\linewidth}
\centering
\begin{tikzpicture}
\begin{axis}[
  ybar,
  width=1.05\linewidth,
    height=0.8\linewidth,
    xlabel=Subsession,
    ylabel=Accuracy (\%),
    ymajorgrids=true,
    xtick align=inside,
  ymin=20, ymax=100,
  xmin=0.5, xmax=10.5,
  xticklabels={1,2,3,4,5,6,7,8,9,10},
  yticklabels={0,20,40,60,80,100},
  x label style={at={(axis description cs:0.5,-0.1)},anchor=north},
  bar width=0.5,
  xtick={1,...,10},enlargelimits=false,
  ytick={0,20,40,60,80,100},clip=false
  ]
  \addplot+[bar shift=0]
  table[x expr=\coordindex+1,y index=1]{\datatable};
  \addplot[bardarkblue, dashed, line legend,sharp plot,nodes near coords={},
update limits=false,shorten >=-3mm,shorten <=-3mm] 
    coordinates {(1.5,90) (9.5,90)}
    node[below,pos=0.05]{$T_{Acc}$};
    \draw[->,myarrow,color=bardarkblue] (2,120) -- (2,110);
    \draw[->,myarrow,color=bardarkblue] (4,120) -- (4,110);
    \draw[->,myarrow,color=bardarkblue] (6,120) -- (6,110);
    \draw[->,myarrow,color=bardarkblue] (8,120) -- (8,110);
    \draw[->,myarrow,color=bardarkblue] (10,120) -- (10,110);
    \node[color=bardarkblue] at (axis cs: -0.5,115) {\textbf{Finetune}};
\end{axis}
\end{tikzpicture}
\caption{\gls{tor} requires adaptation every other subsession, failing to reach the desired accuracy (i.e., five subsessions, 50 training trials of \qty{8.3}{\min}, 56\% accuracy in \qty{16.2}{\second}).}
\label{fig:Example_TOR_subfigA}
\end{subfigure}
~
\begin{subfigure}[t]{0.48\linewidth}
\centering
\begin{tikzpicture}
\begin{axis}[
  ybar,
  width=1.05\linewidth,
    height=0.8\linewidth,
    xlabel=Subsession,
    ylabel=Accuracy (\%),
    ymajorgrids=true,
    xtick align=inside,
  ymin=20, ymax=100,
  xmin=0.5, xmax=10.5,
  xticklabels={1,2,3,4,5,6,7,8,9,10},
  yticklabels={0,20,40,60,80,100},
  x label style={at={(axis description cs:0.5,-0.1)},anchor=north},
  bar width=0.45,
  xtick={1,...,10},enlargelimits=false,
  ytick={0,20,40,60,80,100},clip=false
  ]
  \addplot+[bar shift=0] 
  table[x expr=\coordindex+1,y index=0]{\datatable};
  \addplot[bardarkblue, dashed, line legend,sharp plot,nodes near coords={},
update limits=false,shorten >=-3mm,shorten <=-3mm] 
    coordinates {(1.5,90) (9.5,90)}
    node[below,pos=0.05]{$T_{Acc}$};
    \draw[->,myarrow,color=bardarkblue] (2,120) -- (2,110);
    \node[color=bardarkblue] at (axis cs: -0.5,115) {\textbf{Finetune}};
\end{axis}
\end{tikzpicture}
\caption{\gls{tor} achieves the target threshold after finetuning for one subsession (i.e., one subsession, 10 training trials of \qty{1.6}{\min}, 93.3\% accuracy in \qty{3.2}{\second}).} 
\label{fig:Example_TOR_subfigB}
\end{subfigure}
\caption{Unsuccessful (a) and successful (b) \gls{tor} workflows for $subss=10$ subsessions, $T_{Acc}=90\%$\modified{, $trls=10$ trials.}}
\vspace{-0.5cm}
\label{fig:Example_TOR}
\end{figure}

\begin{figure}[t]
\centering
    \begin{tikzpicture}
    \begin{axis}[
        width=\linewidth,
        height=0.45\linewidth,
        xlabel={Session},
        ylabel={Accuracy (\%)},
        xmin=1.5, xmax=7.5,
        ymin=40, ymax=105,
        xtick={2,3,4,5,6,7},
        ytick={50,60,70,80,90,100},
        x label style={at={(axis description cs:0.5,-0.1)},anchor=north},
        legend cell align={left},
        legend columns = 4,
        legend style={at={(0.45,1.22)},anchor=north,draw=none, font=\scriptsize},
        ymajorgrids=true,
        grid style=dashed,
    ]
    
    \addplot[
        color=newred!110,
        mark=square*,mark options={scale=0.8},
        dashed,
        ]
        coordinates {
        (2,95)
        (3,69.5)
        (4,79)
        (5,87) 
        (6,98) 
        (7,89) 
        };
        \addlegendentry{Chain-\gls{tl}}
    \addplot[
        color=bardarkblue,
        mark=triangle*,mark options={scale=0.8},
        error bars/.cd, y dir=both, y explicit, error bar style={line width=0.1pt, color=black, solid}
        ]
        coordinates {
        (2,76.13) +- (0,16.20)
        (3,61.43) +- (0,21.17)
        (4,55.6) +- (0,10.23)
        (5,81.21) +- (0,16.65)
        (6,93.81) +- (0,6.19) 
        (7,79.14) +- (0,20.05)
        };
        \addlegendentry{TOR-TL}
    \addplot[
        color=bargreen,
        mark=square*,mark options={scale=0.8},
        error bars/.cd, y dir=both, y explicit, error bar style={line width=0.1pt, color=black, solid}
        ]
        coordinates {
        (2,74.83) +- (0,16.53)
        (3,51.6) +- (0,16.17)
        (4,66.30) +- (0,13.92)
        (5,80.0) +- (0,15.90)
        (6,91.75) +- (0,8.25) 
        (7,80.29) +- (0,17.97)
        };
        \addlegendentry{TOR-LwF}  
    \addplot[
        color=barred,
        mark=*,mark options={scale=0.8},
        error bars/.cd, y dir=both, y explicit, error bar style={line width=0.1pt, color=black, solid}
        ]
        coordinates {
        (2,77.19) +- (0,15.05)
        (3,54.8) +- (0,11.36)
        (4,75.67) +- (0,13.83)
        (5,91.32) +- (0,8.68) 
        (6,92.33) +- (0,7.67) 
        (7,82.78) +- (0,17.22) 
        };
        \addlegendentry{TOR-ER}
    \end{axis}
    \end{tikzpicture}
\caption{Test accuracy over multiple sessions for \gls{tl}- and \gls{cl}-based \gls{tor} workflows, and chain-\gls{tl} baseline.}
\label{fig:AvgTestAccuracyonEachSess}
\vspace{-0.3cm}
\end{figure}
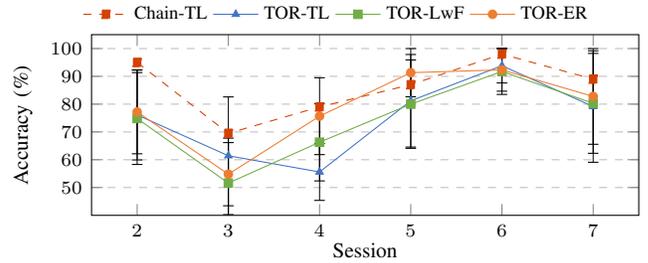

\begin{figure*}[t]
\begin{subfigure}[b]{0.49\textwidth}
\centering
\begin{tikzpicture} 
\begin{axis}[
	width=\linewidth,
    height=0.45\linewidth,
    xmin=1.5, xmax=7.5,
    ymin=0, ymax=85,
    xtick={2,3,4,5,6,7},
    ytick={0,10,20,30,40,50,60,70,80},
	xlabel=Session,
    ylabel=\# of training trials,
	ybar, 
    bar width=3.5pt,
    ymajorgrids=true,
    xtick align=inside,
    x label style={at={(axis description cs:0.5,-0.1)},anchor=north},
    legend cell align={left},
    legend columns = 5,
    legend style={at={(0.45,1.2)},anchor=north,draw=none,fill=none},
    nodes near coords,
    nodes near coords align=vertical,
    nodes near coords style={
    anchor=west, rotate=90, font=\fontsize{6}{8}\selectfont, 
    },
    point meta=explicit symbolic,
]
\addlegendimage{empty legend}
\addlegendentry{\textbf{T\textsubscript{Acc}} \ \ \ }
\addplot[
    color=bardarkblue, fill=bardarkblue, error bars/.cd, y dir=both, y explicit, error bar style={line width=0.25pt, color=black}
    ]
	coordinates {
    (2,20) [76\%] 
    (3,48) [51\%] 
	(4,26) [70\%] 
    (5,16) [80\%] 
    (6,10) [91\%] 
	(7,16) [75\%] }; 
 \addlegendentry{70\%}
\addplot[
    color=bargreen, fill=bargreen, error bars/.cd, y dir=both, y explicit, error bar style={line width=0.25pt, color=black}
    ]
	coordinates {
    (2,24) [78\%] 
    (3,50) [54\%] 
	(4,36) [71\%] 
    (5,26) [83\%] 
    (6,8) [91\%] 
	(7,16) [82\%]}; 
  \addlegendentry{80\%}
\addplot[
    color=baryellow, fill=baryellow, error bars/.cd, y dir=both, y explicit, error bar style={line width=0.25pt, color=black}
    ]
	coordinates {
    (2,36) [77\%] 
    (3,50) [55\%] 
	(4,40) [76\%]
    (5,24) [91\%] 
    (6,14) [92\%] 
	(7,28) [83\%]}; 
  \addlegendentry{90\%}
\addplot[
    color=barred, fill=barred, error bars/.cd, y dir=both, y explicit, error bar style={line width=0.25pt, color=black}
    ]
	coordinates {
    (2,46) [77\%] 
    (3,50) [58\%] 
	(4,40) [77\%] 
    (5,34) [90\%] 
    (6,22) [92\%] 
	(7,36) [84\%]}; 
  \addlegendentry{100\%}
\addplot[
    color=newred!110, fill=newred!110, error bars/.cd, y dir=both, y explicit, error bar style={line width=0.25pt, color=black}
    ]
	coordinates {
    (2,60) [95\%]
    (3,60) [70\%]
	(4,60) [79\%]
    (5,60) [87\%]
    (6,60) [98\%]
	(7,60) [89\%]}; 
\end{axis}
\end{tikzpicture}
\caption{\modified{We consider $T_{Acc} \in \{70\%, 80\%, 90\%, 100\%\}$ for $trls=10$.}}
\label{fig:AblationStudies_subfigA}
\end{subfigure}
\hfill
\begin{subfigure}[b]{0.49\textwidth}
\centering
\begin{tikzpicture} 
\begin{axis}[
	width=\linewidth,
    height=0.45\linewidth,
    xmin=1.5, xmax=7.5,
    ymin=0, ymax=85,
    xtick={2,3,4,5,6,7},
    ytick={0,10,20,30,40,50,60,70,80},
	xlabel=Session,
    ylabel=\# of training trials,
	ybar, 
    bar width=4.5pt,
    ymajorgrids=true,
    xtick align=inside,
    x label style={at={(axis description cs:0.5,-0.1)},anchor=north},
    legend cell align={left},
    legend columns = 4,
    legend style={at={(0.45,1.2)},anchor=north,draw=none,fill=none},
    nodes near coords,
    nodes near coords align=vertical,
    nodes near coords style={
    anchor=west, rotate=90, font=\fontsize{6}{8}\selectfont, 
    },
    point meta=explicit symbolic,
]
\addlegendimage{empty legend}
\addlegendentry{\textbf{trls} \ \ \ }
\addplot[
    color=bardarkblue, fill=bardarkblue, error bars/.cd, y dir=both, y explicit, error bar style={line width=0.25pt, color=black}
    ]
	coordinates {
    (2,39) [82\%] 
    (3,45) [63\%] 
	(4,46) [65\%] 
    (5,32) [86\%] 
    (6,23) [93\%] 
	(7,33) [84\%] }; 
\addlegendentry{5}
\addplot[
    color=bargreen, fill=bargreen, error bars/.cd, y dir=both, y explicit, error bar style={line width=0.25pt, color=black}
    ]
	coordinates {
    (2,36) [77\%] 
    (3,50) [55\%] 
	(4,40) [76\%] 
    (5,24) [91\%] 
    (6,14) [92\%] 
	(7,28) [83\%] }; 
\addlegendentry{10}
\addplot[
    color=barred, fill=barred, error bars/.cd, y dir=both, y explicit, error bar style={line width=0.25pt, color=black}
    ]
	coordinates {
    (2,40) [71\%] 
    (3,40) [48\%] 
	(4,40) [66\%] 
    (5,32) [86\%] 
    (6,20) [89\%] 
	(7,32) [78\%] }; 
\addlegendentry{20}
\addplot[
    color=newred!110, fill=newred!110, error bars/.cd, y dir=both, y explicit, error bar style={line width=0.25pt, color=black}
    ]
	coordinates {
    (2,60) [95\%]
    (3,60) [70\%]
	(4,60) [79\%]
    (5,60) [87\%]
    (6,60) [98\%]
	(7,60) [89\%]}; 
\end{axis}
\end{tikzpicture}
\caption{\modified{We consider $trls \in \{5, 10, 20\}$ for $T_{Acc}=90\%$.}}
\label{fig:AblationStudies_subfigB}
\end{subfigure}
\caption{Total number of training trials and test accuracy for \gls{tor}-\gls{er} varying the accuracy threshold $T_{Acc}$ (a) and the number of trials in a subsession $trls$ (b). Chain-\gls{tl} baseline is represented in red.}
\label{fig:AblationStudies}
\vspace{-0.3cm}
\end{figure*}
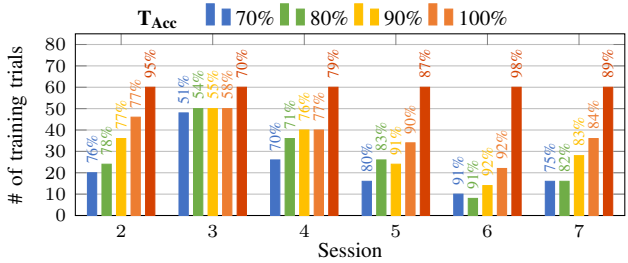
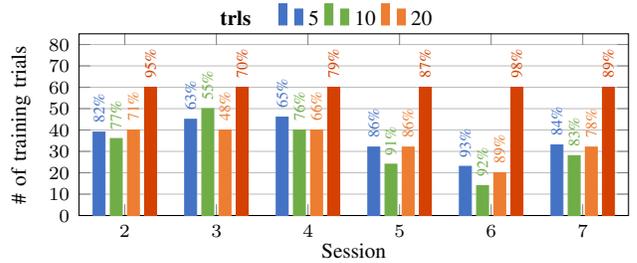

In~\Cref{fig:NumOfTrainingTrials}, we analyze the total amount of required training trials for three learning techniques: \gls{tl}-, \gls{er}-, and \gls{lwf}-based \gls{tor}, compared with a chain-\gls{tl} baseline.  
\modified{Notably, our proposed \gls{tor} strategy requires fewer training samples than chain-\gls{tl} baseline. 
As the number of \gls{tor} training trials in each session varies, we depict the worst-case (i.e., session 3) and the best-case (i.e., session 6) scenarios in~\Cref{fig:Example_TOR}.
\Cref{fig:Example_TOR_subfigA} shows an out-of-distribution session, where the finetuned model fails to reach the desired accuracy threshold throughout the session. 
Conversely, ~\Cref{fig:Example_TOR_subfigB} shows a successful scenario on in-distribution data, where \gls{tor} improves the accuracy (i.e., up to 100\% within a subsession) while reducing the number of training samples (i.e., only \qty{1.6}{\minute} of acquisition over the entire session) compared to the baseline.}

When extending the intra-session analysis to an inter-session study \modified{in~\Cref{fig:NumOfTrainingTrials}}, we note that the \modified{amount of data needed to achieve satisfactory performance} decreases with time, demonstrating the learning potential of our \gls{tor} workflow for long-term usage of \gls{bmi} systems.
Moreover, while chain-\gls{tl} requires 60 training trials, the proposed \gls{er}-based \gls{tor} achieves satisfactory accuracy levels (i.e., up to 92.33\% \modified{average test accuracy} on session 6) using 46.67\% fewer data samples.
This results in a total acquisition time of \qty{32}{\min}, i.e., \qty{5.3}{\minute} per session on average, for 192 trials recorded over six sessions, averaged over five runs. 
\modified{This is 9\% (\qty{3}{\min}) fewer than \gls{tl}-based \gls{tor} and 46\% (\qty{28}{\min}) fewer than chain-\gls{tl} baseline.}

In~\Cref{fig:AvgTestAccuracyonEachSess}, we further analyze the proposed \gls{tor} workflows considering the attainable classification accuracy per session.
We report the test accuracy averaged over the tested subsessions within a session.
When comparing the \gls{tl}- and \gls{cl}-based \gls{tor} approaches, \gls{er} achieves the highest average accuracy of 79\% over six sessions, 4.47\% over \gls{tl}, while requiring the least amount of training data.
We additionally note that the accuracy increases with time, reaching an average of 88.81\% in the last three sessions, which is comparable with chain-\gls{tl} accuracy, showing that our methodology enables the MI-BMInet backbone to gain and preserve knowledge over time. Moreover, the average \gls{itr}~\cite{frey2024gapses} in the last three sessions is \qty[per-mode = symbol]{9.0}{\bit \per \minute} for chain-TL and \qty[per-mode = symbol]{7.6}{\bit \per \minute} for \gls{tor}-\gls{er}.

\subsection{Sensitivity Analysis on \gls{tor} Constraints}

Our proposed \gls{tor} workflow depends on the trial granularity within a session and on the accuracy threshold that decides the finetuning process.
We measure their impact on the test accuracy and on the training trials required \modified{to achieve a satisfactory accuracy}.
We first fix $trls=10$ trials in a subsession, while sweeping $T_{Acc}\in\{70\%, 80\%, 90\%, 100\%\}$ with \gls{er}-based \gls{tor}.
When the threshold increases, so does the amount of data required, shown in~\Cref{fig:AblationStudies_subfigA}, as more on-site information is needed to capture the class distribution in the target domain.
As depicted in~\Cref{fig:Example_TOR_subfigA}, imposed thresholds are not always reachable, yet with higher thresholds and more training data, we notice accuracy levels up to 58\% in complex sessions and 92\% in in-distribution settings.

We also analyse the number of required training trials and the achieved test accuracy under \gls{er}-based \gls{tor}, sweeping $trls \in \{5,10,20\}$ while keeping $T_{Acc}=90\%$.
While a coarse-grained division of trials within a subsession enables the model to access more information over the target distribution, a fine-grained division speeds up the testing and training processes. 
As shown in~\Cref{fig:AblationStudies_subfigB}, using subsessions of 10 trials ensures an optimal accuracy-training time trade-off, with an average test accuracy of 79\% and an average of \qty{5.3}{\minute} of acquisition time.
Note that these parameters can be set by the users according to their acceptance of accuracy and calibration time trade-off.

\subsection{On-Device Learning Measurements}

Tailored for extreme edge \glspl{mcu}, we deploy MI-BMInet and implement our \gls{odl} routine on the GAP9 processor. 
Using a RISC-V \gls{fc} core as the system controller, GAP9 can delegate computational tasks to an 8-core RISC-V compute cluster, enabling efficient parallelization of the \gls{dl} workload with a power efficiency of \qty{0.33}{\milli \watt /GOP}~\cite{GreenWavesGAP9}.
Four shared \glspl{fpu} are present in the computational cluster.
Moreover, GAP9 features a shared L1 \gls{tcdm} memory of \qty{128}{\kilo \byte}, enabling single-cycle access from the computational cores.
Additional \qty{1.6}{\mega \byte} L2 RAM memory and \qty{2}{\mega \byte} L3 non-volatile eMRAM are present at \gls{fc} level for the deployment of \glspl{dnn} and storage of training trials.

We freeze and quantize our MI-BMInet backbone in \texttt{int8} and deploy it using Dory~\cite{burrello2020dory}, whilst the precision of the retrainable classifier is maintained to \texttt{fp32}.
We thus benefit from the \gls{simd} instructions for the integerized backbone employed only for the forward pass, whilst the backward pass and the gradient computation can be performed without loss of accuracy in single-precision floating point. 
We implement our proposed \gls{odl} workflow using PULP-TrainLib \cite{NadaliniPulpTrainLib2022} learning library targeting PULP~\glspl{mcu}. 
By training the last linear layer on GAP9, we require only \qty{1.08}{\milli \joule} for a runtime of \qty{21.6}{\milli \second} per training step using one data trial, i.e., an average power consumption of \qty{50.2}{\milli \watt}\modified{, operating at \qty{800}{\milli \volt} and \qty{370}{\mega \hertz}}.
\modified{Compared to the acquisition time of \qty{1.6}{\minute} per subsession, \gls{tor} on-board training for 15 epochs takes only \qty{3.2}{\second} per subsession, enabling on-site adaptation without sacrificing user experience.}

\section{Conclusion}
\label{sec:conclusion}

This work proposes a novel \gls{tor} \gls{odl} workflow aimed for \gls{eeg}-based \gls{bmi} systems, which addresses inter-session variability in users through near real-time data acquisition and online finetuning to out-of-distribution sessions while also significantly reducing the training time.
Addressing catastrophic forgetting, we enable adaptation to novel conditions by integrating \gls{er}-based learning in our proposed \gls{tor} workflow.
\modified{We thus achieve a 46\% calibration time reduction down to \qty{1.6}{\minute} compared to the baseline, while reaching accuracy levels up to 92\%.}
Furthermore, we deployed our \gls{odl} strategy on GAP9 \gls{mcu}, showing a power consumption of \qty{50}{\milli \watt} and demonstrating a modest learning cost of \qty{1}{\milli \joule} per training step \modified{within a short runtime of \qty{21.6}{\milli \second}.} 
By providing a robust, user-friendly, and energy-efficient implementation, these results bring us closer to achieving full user acceptance of \glspl{bmi}, even in challenging non-clinical scenarios.

\section*{Acknowledgment}
This project was supported by the Swiss National Science Foundation under grant agreement 193813 (Project PEDESITE) and grant agreement 207913 (Project TinyTrainer), by the ETH-Domain Joint Initiative program (project UrbanTwin), and by the ETH Future Computing Laboratory (EFCL). 

\bibliographystyle{IEEEtran}
\bibliography{bib}

\end{document}